\documentstyle[twoside]{article}
\def\runninghead#1#2{\pagestyle{myheadings}
\markboth{\hfill{\protect\footnotesize\it{\quad #1}}}
{{\protect\footnotesize\it{#2\quad}}\hfill}}
\begin{document}

\runninghead{\bf Chubykalo  et al.}
{\bf Reply to ``Comment ..."}

$$$$
{\large {\bf REPLY TO `COMMENT ON ``HELMHOLTZ THEOREM AND THE V-GAUGE IN THE PROBLEM OF SUPERLUMINAL AND INSTANTANEOUS SIGNALS IN CLASSICAL ELECTRODYNAMICS" BY A. CHUBYKALO ET AL'  BY J. A. HERAS [FOUND. PHYS. LETT. vol. 19(6) p. 579 (2006)]}
\bigskip
\bigskip

$\;\;\;\;\;\;\;\;\;\;${\bf A. Chubykalo, A. Espinoza, R. Alvarado Flores$^a$

$\;\;\;\;\;\;\;\;\;\;$and A. Gutierrez Rodriguez}

\bigskip

$\;\;\;\;\;\;\;\;\;\;${\it Escuela de F\'{\i}sica, Cuerpo Academico ``Particulas, campos

$\;\;\;\;\;\;\;\;\;\;$y astrof\'{\i}sica", Universidad Aut\'onoma
de Zacatecas,

$\;\;\;\;\;\;\;\;\;\;$Apartado Postal C-580\, Zacatecas 98068, ZAC., M\'exico

$\;\;\;\;\;\;\;\;\;\;$e-mail: achubykalo@yahoo.com.mx}

\bigskip

$\;\;\;\;\;\;\;\;\;\;^a\;${\it Centro de estudios multidisciplinarios,

$\;\;\;\;\;\;\;\;\;\;$Universidad Aut\'onoma de Zacatecas}

\bigskip

$\;\;\;\;\;\;\;\;\;\;$Received $\;\;\;\;\;\;\;\;\;\;\;\;$, 2007

\baselineskip 5mm

\bigskip

\bigskip
\noindent
Key words: Helmholtz theorem, v-gauge, electromagnetic potentials, electromagnetic waves.

$$$$

\bigskip
\noindent Jos\'{e} A. Heras has recently raised interesting
criticism [1] to the conclusions to our paper [2]. His main point
is his assertion that the equations we derived for the solenoidal
(${\bf E}_s$) and the irrotational (${\bf E}_i$) components of the
electric field ${\bf E}$ are a  pair of coupled equations that can
be reduced to one equation.  If this is the case, our claim that
there exists two mechanisms of energy and momentum propagation is
incorrect and our example using a bounded oscillating charge on
the $x$-axis must be incorrect and Heras attempts to find
contradictions in our treatment. Therefore in this paper we will
show that the equations for ${\bf E}_i$ and ${\bf E}_s$ are not
coupled equations. We indicate the kind of problems that classical
electrodynamics in its standard form cannot neither explain nor
predict, however, such problems can be solved using Helmholtz
theorem and our interpretation of its use. Hence our claim that
there exist two mechanisms of energy and momentum transfer, is, to
our knowledge and according to our interpretation, correct.

\bigskip

Now we  show that the equations for the
solenoidal and the irrotational components of the electric field are not coupled
equations.

So, we start from standard Maxwell's equations

\begin{eqnarray}
\nabla \times {\bf B} &=&\frac{4\pi}{c}{\bf j}+\frac 1c\frac{\partial
{\bf E}}{\partial t}\;,   \\
\nabla \times {\bf E} &=&-\frac{1}{c}\frac{\partial {\bf B}}{\partial t}\;,
 \\
\nabla \cdot {\bf E} &=&4\pi \varrho\;,    \\
\nabla \cdot {\bf B} &=&0\;.
\end{eqnarray}

 From  (1)-(3)
we can get the following wave equation

\begin{equation}
\nabla^2{\bf E}-\frac{1}{c^2}\frac{\partial^2{\bf E}}{\partial
t^2}= 4\pi\left(\nabla\varrho +\frac{1}{c^2}\frac{\partial{\bf
j}}{\partial t} \right).
\end{equation}
If we apply Helmholtz
theorem and equation (4) we obtain three more equations

\begin{equation}
{\bf E} ={\bf E}_i+{\bf E}_s, \qquad{\bf j} ={\bf j}_i+{\bf j}_s,\qquad {\bf B} ={\bf B}_s.
\end{equation}
Using (6) the Maxwell's equations become

\begin{eqnarray}
\nabla \times {\bf B}_s &=&\frac{4\pi}{c}{\bf j}_s+\frac{4\pi}{c}{\bf j}_i+\frac {1}{c}\frac{\partial {\bf E}_s}{\partial t}+\frac{1}{c}\frac{\partial
{\bf E}_i}{\partial t}\;,   \\
\nabla \times {\bf E}_s &=&-\frac {1}{c}\frac{\partial {\bf B}_s}{\partial
t}\;,   \\
\nabla \cdot {\bf E}_i &=&4\pi \varrho\;,    \\
\nabla \cdot {\bf B}_s &=&0\;.
\end{eqnarray}
A time differentiation of (7) and the use of (8) give us the equation

\begin{equation}
\nabla^2{\bf E}_s-\frac{1}{c^2}\frac{\partial^2{\bf E}_s}{\partial
t^2}=\frac{4\pi}{c^2}\frac{\partial {\bf j}_s}{\partial t}+\frac{4\pi }{c^2}\frac{\partial {\bf j}_i}{\partial t}+\frac{1}{c^2}
\frac{\partial ^2{\bf E}_i}{\partial t^2}.
\end{equation}
Equation (11) is Heras' key equation. In this equation it seems that the
solenoidal and the irrotational components of the electric field are
coupled. If equations (7) and (8) were our only equations at hand
undoubtedly Heras would be right. However  we have equation
(9). This equation is independent of equations (7) and (8), so we can use
it to decouple the irrotational and solenoidal components in equation (11).

We can prove [2] the equation

\begin{equation}
-\frac{\partial {\bf E}_i}{\partial t}=4\pi {\bf j}_i
\end{equation}
from equation (9) directly using the continuity equation, Helmholtz theorem and the asymptotic behavior of harmonic functions, as follows:

$$
{\bf E}_i= -\nabla\int\frac{\nabla^{\prime}\cdot{\bf E}({\bf r}^{\prime},t)}{4\pi|{\bf r}-{\bf r}^{\prime}|}dV^{\prime}=
-\nabla\int\frac{\varrho({\bf r}^{\prime},t)}{|{\bf r}-{\bf r}^{\prime}|}dV^{\prime}\Rightarrow
$$

$$
\frac{\partial {\bf E}_i}{\partial t}=-\nabla\int\frac{\frac{\partial \varrho({\bf r}^{\prime},t)}{\partial t}}{|{\bf r}-{\bf r}^{\prime}|}dV^{\prime}=
\nabla\int\frac{\nabla^{\prime}\cdot{\bf j}({\bf r}^{\prime},t)}{|{\bf r}-{\bf r}^{\prime}|}dV^{\prime}=-4\pi {\bf j}_i
$$

If we combine equations (11) and
(12) we get for ${\bf E}_s$

\begin{equation}
\nabla^2{\bf E}_s-\frac{1}{c^2}\frac{\partial^2{\bf E}_s}{\partial
t^2}=\frac{4\pi }{c^2}\frac{\partial \mathbf{j}_s}{\partial t}.
\end{equation}
For ${\bf E}_i$ we must get a differential equation too. To do it
we use Eqs. (6)  at Eq. (5), then we  get for the irrotational
component the equation

\begin{equation}
\bigtriangleup {\bf E}_i=4\pi \nabla \varrho\;.
\end{equation}

Equations (13) and (14) are the basic equations of our paper [2],
and they are a direct consequence of Maxwell's equations and
Helmholtz theorem.

It is no correct to claim that equation (14) follows from
equations (1) and (5) follows [1].  We have seen that equation
(14) follows from equation (11), the continuity equation, and the
asymptotic behavior of harmonic functions at spatial infinity.
Indeed, we must remember that the term $\frac{\partial {\bf
E}}{\partial t}$ was included by Maxwell himself on the right hand
side of equation (1) to avoid any contradiction with the
continuity equation, so, Maxwell considered the charge
conservation as a fundamental law that must be satified by
Amp\`ere law. In this sense, equation (14) is independent of
equation (1) because we need not equation (1) to deduce equation
(14). Therefore we have proved that our basic equations (13) and
(14) are independent, so the irrotational and the solenoidal
components of the electric field are determined {\it by
independent} differential equations. This is {\it the most
important achievement} of our paper [2].

However we must point out that Heras' criticism [1] is, as it
were, twofold: a mathematical technique that supports a physical situation.
 When he says that the components of the electric field are coupled the underlying
  physical intuition and philosophical commitment is that in fact the irrotational component
   is no detectable and cannot mediate the electromagnetic interaction. So, in some sense, the
   non-physical components must be related to the physical ones and this relation must be expressed
    mathematically. For this reason the supposition that Eq.(11) that relates both components is a natural one, {\it however}, our critical analysis shows {\it that this is not the case}: both components satisfy {\it different} uncoupled equations. Now we must show that a simple physical situation exists  where the components may have a separate physical meaning.

We must show now that both fields can be realized as independent physical quantities. First of all, we point out in detail a problem that classical electrodynamics in its standard form cannot solve.

We start with the following question:  Is it possible for the retarded
solutions of the wave equation to explain all electromagnetic phenomena? We
claim that this is not the case, as the following simple example (which is
no other than the one used in [2]) shows.

Consider a point charge at rest at the origin of an inertial
cartesian coordinate system. If it begins to move along the
$x$-axis there is a brief interval of time when the charge is
accelerated, and the electromagnetic field of the moving charge is
given by the retarded fields ${\bf E}_{ret},{\bf B}
_{ret}=\frac{{\bf R}}{R}\times {\bf E}_{ret}$ where ${\bf R}$ is
the position vector of the particle, in this case along the
$x$-axis. The question is now: {\it Is there energy and momentum
propagation along the $x$-axis?} Because along this axis the
velocity ${\bf V}$ and  the acceleration ${\bf a}$ of the particle
are parallel to ${\bf R}$, the acceleration field is clearly zero.
Hence in the retarded electric field only the velocity field
survives

\begin{equation}
{\bf E}_{ret}=\frac{q\left( \frac{{\bf R}}R-\frac{{\bf V}}{c}\right)
}{\gamma ^2\left( 1-\frac{{\bf R\cdot V}}{Rc}\right) R^2}\mid _{ret}\;,
\end{equation}
so we obtain for the retarded magnetic field ${\bf
B}_{ret}=\frac{{\bf R}}{R}\times{\bf E}_{ret}=0$.

Along the $x$-axis the Poynting vector ${\bf S}=0$. Therefore along the $x$-axis
there is no transmission of energy and momentum, and any test charge in this
axis does not feel the presence of the moving charge initially at rest at
the origin. But this contradicts experience, because if the charge initially at
 rest begins to move, any test charge along the $x$-axis begins to move too, so, a transmission of energy and
  momentum from one charge to the other {\it intervenes}! How comes it? Obviously trough the Poynting vector
   this is not possible because there is not energy flow along the   $x$-axis. Hence the obvious logical conclusion is
    that there must be other mechanism of energy transmission from an accelerated charge to other charges. In our
     original paper [2] we used an example where ${\bf E}_s$ is zero along the line of action, and for this reason it
      is not possible to find any contradiction with the retarded solutions.

Now that our motivation is clear we must return to Heras'
arguments. He claims that when ${\bf E}={\bf E}_i$ Faraday's law
is violated. For our case it is clear that ${\bf B}_s=0$, so  any
violation of Faraday's law is not possible. However, even in
general terms, it is not the case that when the electric field is
purely irrotational Faraday's law  is violated, because when we
write  Maxwell's equations in solenoidal form,  Faraday's law
becomes

$$
\nabla \times {\bf E}_s=-\frac{1}{c}\frac{\partial {\bf B}_s}{\partial t}
$$
as can be seen from equation (8). Now, if ${\bf B}_s=0$ the vector
field becomes purely irrotational as can be deduced from equations
(7-10), and if ${\bf E}_s=0$ then any time-independent magnetic
field would satisfy Faraday's law. So, from Faraday's law Heras
cannot obtain any contradiction.

We have expounded our motivation for the use of Helmholtz theorem:
it is necessary to explain energy and momentum transmission in
some simple cases where the retarded solutions are impotent, and
we believe that the use of  Helmholtz theorem is a way to achieve
this goal using standard Maxwell's equations. We have replied to
Heras too, showing that his supposed violation of Faraday's law is
a misunderstanding. As a conclusion we still claim that there are
two mechanisms of transmission.

In this reply we have shown in detail that the criticism raised by J. A.
Heras to our paper [2] are the result of a misunderstanding of the
fundamentals of our work. For that reason  we explained the
kind of problems that motivate our theory of the two mechanisms of
trasmission of energy and momentum and  we deduced, once again
and with logical detail, the basic equations of our theory.
 To summarize one can say that our theory is quite simple:  we have two
  {\it independent} equations for the irrotational and solenoidal
   components of the electric field, {\it but this is not enough for
    a physical theory}; that would be  a point in favor of Heras; and for this
     reason, we must show that there are physical situations where the influence
      of the electric field can be reduced to the influence of its irrotational component.
       We think that we have achieved this goal in our paper [2].

$$$$ {\large {\bf REFERENCES}}

\end{document}